# Relaxed current matching requirements in highly luminescent perovskite tandem solar cells and their fundamental efficiency limits

Alan R. Bowman[1], Felix Lang[1], Yu-Hsien Chiang[1], Alberto Jiménez-Solano[2], Kyle Frohna[1], Giles E. Eperon[3], Edoardo Ruggeri[1], Mojtaba Abdi-Jalebi[1,†], Miguel Anaya[1], Bettina V. Lotsch[2,4,5] & Samuel D. Stranks[1,6]*

1. Cavendish Laboratory, Department of Physics, University of Cambridge, JJ Thomson Avenue, Cambridge, CB3 0HE, United Kingdom
2. Max Planck Institute for Solid State Research, Nanochemistry Department, Heisenberg Str. 1, 70569, Stuttgart, Germany
3. National Renewable Energy Laboratory, 16253 Denver West Parkway, Golden, CO 80401, United States of America
4. Department of Chemistry, Ludwig-Maximilians-Universität (LMU), Butenandtstrasse 5-13, 81377 Munich, Germany
5. E-conversion Lichtenbergstrasse 4a, 85748, Garching, 80799 Munich, Germany
6. Department of Chemical Engineering & Biotechnology, University of Cambridge, Philippa Fawcett Drive, Cambridge, CB3 0AS, United Kingdom
†Current address: Institute for Materials Discovery, University College London, Torrington Place, London, WC1E 7JE, United Kingdom
*Corresponding author: sds65@cam.ac.uk

**Abstract**

Perovskite based tandem solar cells are of increasing interest as they approach commercialisation. Here we use time-resolved and steady-state optical spectroscopy on state-of-the-art low- and high-bandgap perovskite films for tandems to quantify intrinsic recombination rates and absorption coefficients. We apply these experimental parameterised data to calculate the limiting efficiency of perovskite-silicon and all-perovskite two-terminal tandems employing currently available bandgap materials as 42.0 % and 40.8 % respectively. By including luminescence coupling between sub-cells, i.e. the re-emission of photons from the high-bandgap sub-cell and their absorption in the low-bandgap sub-cell, we reveal the stringent need for current matching is relaxed when the high-bandgap sub-cell is a luminescent perovskite compared to calculations that do not consider luminescence coupling. We show that luminescence coupling becomes important in all-perovskite tandems when charge carrier trapping rates are $< 10^6$ s$^{-1}$ (corresponding to carrier lifetimes longer than 1 µs at low excitation densities) in the high-bandgap sub-cell, which is lowered to $10^5$ s$^{-1}$ in the better-bandgap-matched perovskite-silicon cells. In both tandem technologies, this threshold corresponds to a high-bandgap sub-cell with an external luminescence quantum efficiency of at least ~0.1 % at maximum power point. We demonstrate luminescence coupling endows greater flexibility in both sub-cell thicknesses, increased tolerance to different spectral conditions and a reduction in the total thickness of light absorbing layers. To maximally exploit luminescence coupling we reveal a key design rule for luminescent perovskite-based tandems: the high-bandgap sub-cell should always have the higher short-circuit current. Importantly, this can be achieved by reducing the bandgap or increasing the thickness in the high-bandgap sub-cell with minimal reduction in efficiency, thus allowing for wider, unstable bandgap compositions (>1.7 eV) to be avoided. Finally, we experimentally visualise luminescence coupling in an all-perovskite tandem device stack through cross-section luminescence images, revealing that the effect must already be considered for further perovskite tandem development.

**Introduction**

The performances of halide perovskite solar cells, epitomised by the workhorse methylammonium lead iodide (MAPbI$_3$) composition, have rapidly improved over the last decade and power conversion efficiencies now rival those of silicon[1]. Perovskites are ideal light harvesting layers for solar cells due to strong absorption coefficients, long charge diffusion lengths and tolerance to charge traps[2]. Importantly, the bandgap of halide perovskites can be controlled via the substitution of a fraction of lead for tin (lowering the bandgap from ~1.6 eV to ~1.2 eV) or chlorine and bromine for iodine (raising the bandgap to ~ 2.3 eV and 3 eV respectively in pure-lead systems)[3,4]. This tunability means perovskites hold great promise for realising cheap and efficient tandem solar cells in which two absorber layers of different bandgaps harvest complementary regions of the solar spectrum. To date, all-perovskite tandems have achieved efficiencies of 24.8 % and silicon-perovskite tandems of 29.1 %[1,5]. Importantly, both of these tandem technologies are predicted to realise low enough levelised cost of electricity to make them competitive with market-leading single bandgap silicon solar cells[6]. As tandem perovskite solar cells continue to improve it is important to understand fully their thermodynamic efficiency limits and any current-matching conditions required for optimal operation, both of which impose restrictions on material and device design. While there are several reports estimating all-perovskite and perovskite-silicon tandem efficiency limits and optimal optical designs, the majority focus on what is achievable with current technologies (e.g. for transmission from top contacts) and, critically, do not include all intrinsic recombination and luminescence coupling processes[7–13]. This means that tandem device optimisation is currently being guided by incomplete models that do not capture all effects.

Here we measure intrinsic recombination rates and absorption coefficients in perovskite thin films using time-resolved and steady-state optical spectroscopy, and use these values to calculate the thermodynamic efficiency limit of low-bandgap perovskite formamidinium lead-tin iodide (FAPb$_{0.5}$Sn$_{0.5}$I$_3$) as 32.1 %, an all-perovskite tandem, with the same low bandgap system coupled to FA$_{0.7}$Cs$_{0.3}$Pb(I$_{0.7}$Br$_{0.3}$)$_3$, as 40.8 %, and this high-bandgap system coupled to an idealised silicon absorber layer as 42.0 % (using literature recombination rates and absorption coefficients for silicon). We demonstrate that consideration of luminescence coupling between sub-cells, i.e., the emission of light from the high-bandgap sub-cell and its subsequent reabsorption in the low-bandgap sub-cell, relaxes the need for current matching compared to previous calculations that do not include the effect. It becomes important when charge trapping rates are <10$^6$ s$^{-1}$ in all-perovskite tandems and <10$^5$ s$^{-1}$ in perovskite-silicon tandems – values comparable to what are already achievable in reported materials, where charge lifetimes in the charge-trapping regime are on the order of 1-10 µs[14,15]. Furthermore, by exploring a range of experimental device optimisation parameters, consideration of luminescence coupling allows for greater flexibility in the choice of sub-cell thicknesses and bandgaps in a tandem cell, alongside increased tolerance to a range of real-world spectral conditions. Using an all-perovskite tandem cell, we provide proof-of-concept spectroscopic visualisation and electrical measurements of luminescence coupling, demonstrating the direct implications of our work for further perovskite device optimisation.

**Results and Discussion**

The maximum efficiency of a single bandgap solar cell was derived in the seminal paper by Shockley and Queisser[16] and extended to include any number of ideal tandem solar cells by de Vos[17]. Considerations relevant to specific material systems, for example non-ideal absorption and intrinsic non-radiative loss mechanisms, were first included for single junction silicon solar cells, and more recently for MAPbI$_3$ solar cells[18,19]. Efficiency models are based on calculating the extracted current as the difference between generated charges, $J_{sc}$, and those lost to recombination, $J_0(V)$:

$$J(V) = J_{sc} - J_0(V). \tag{1}$$

Here $V$ is the voltage across the semiconductor (in an ideal case assumed equal to the Fermi-level splitting), and the maximum efficiency is found by maximising the product $\frac{JV}{incident\ power}$.

When considering tandem solar cells, an additional intrinsic process should be included when compared to single junction devices: luminescence coupling between the sub-cells. This phenomenon has been previously explored in idealised systems[20,21] and III-V tandem technologies[22–28], but has not been considered in perovskite technologies. We first briefly discuss luminescence coupling, especially its importance to perovskite tandems, within an idealised Shockley-Queisser formalism, before presenting results using experimental parameters and including other non-ideal absorption and loss processes. We consider that the density of black body radiation is higher in a semiconductor than in its surroundings by a factor of $n(E)^2$, the real refractive index at energy $E$, due to the increased density of states[29]. Light emitted from any material can only interact with its surroundings (where $n(E)\sim 1$) through its light escape cone, reducing the fraction of black body radiation by a factor of $\frac{1}{n(E)^2}$ (so semiconductors are still in equilibrium with their surroundings). However, between two tandem sub-cell absorbers with refractive indices larger than 1, the escape cone covers a solid angle a factor of $n(E)_x^2$ larger than with the surroundings (where $x$ refers to the lowest index of refraction of the semiconductors). Typically, the high-bandgap sub-cell can only absorb a small fraction of the light emitted by the low-bandgap sub-cell, as most emitted light is below its bandgap. However, the low-bandgap cell can absorb a significant fraction of radiation emitted by the high-bandgap cell (see schematic in inset of Figure 1b). In a two-terminal tandem solar cell the same current must flow through both sub-cells. Therefore, at maximum power point the maximum number of extracted charges is determined by the sub-cell with the minimum number of photo-generated charges. If the low-bandgap sub-cell is the limiting sub-cell, charges not extracted from the high-bandgap sub-cell can recombine radiatively and be re-absorbed in the low-bandgap sub-cell, reducing the current mismatch.

In Figure 1, we present the limiting efficiency of Shockley-Queisser-like tandem solar cells under AM1.5 (where all light above the bandgap is absorbed and the only loss process is radiative recombination) without considering light coupling between the sub-cells (Figure 1a), as has been presented in the perovskite field to date, and compare this to the case including luminescence coupling (Figure 1b). The ratio of these efficiencies is shown in Figure 1c. We note it is possible to prevent luminescence coupling in this fully idealised case by use of a suitable dichroic mirror between the sub-cells (reflecting all light emitted from the back of the high-bandgap sub-cell), while in real systems it cannot be prevented due to absorption coefficients not being step functions (i.e. there will be a spectral region where both high- and low-bandgap sub-cells absorb light and can therefore couple). Here we have used the value of $n(E)_x = 2.5$, representative of metal halide perovskites (cf. Figure 2c, d). Luminescence coupling between layers lowers the maximum possible efficiency from 45.8 % to 44.9 % due to more light being lost from the high-bandgap sub-cell than in the case without coupling. While the optimal bandgap pair remains within 0.01 eV of that without luminescence coupling (0.94 eV and 1.60 eV for the low- and high-bandgap sub-cells respectively), Figure 1b demonstrates that luminescence coupling gives greater tolerance in the choice of sub-cell bandgaps to achieve a high efficiency. Specifically, when the high-bandgap sub-cell has a significantly larger short circuit current than the low-bandgap sub-cell, efficiency is increased when luminescence coupling is included. This beneficial region can be seen below the diagonal dashed line of Figure 1c (see SI Figure S1 for a plot of the short circuit current in each sub-cell and a ratio of the two). To further illustrate how this result impacts device design for the case of halide perovskites, we plot line-slices of Figure 1a and 1b in Figure 1d with the low bandgap fixed at 1.25 eV, close to the lowest bandgap currently technically feasible for halide perovskites[30], and we vary the high bandgap. Our results demonstrate that the high bandgap can be reduced to as low as ~1.6-1.7 eV with minimal loss in efficiency, compared to the much less stable bandgaps in the 1.8-1.9 eV range (which typically require high fractions of bromide and/or caesium) required in the case without luminescence coupling[7], by relaxing current-matching requirements.

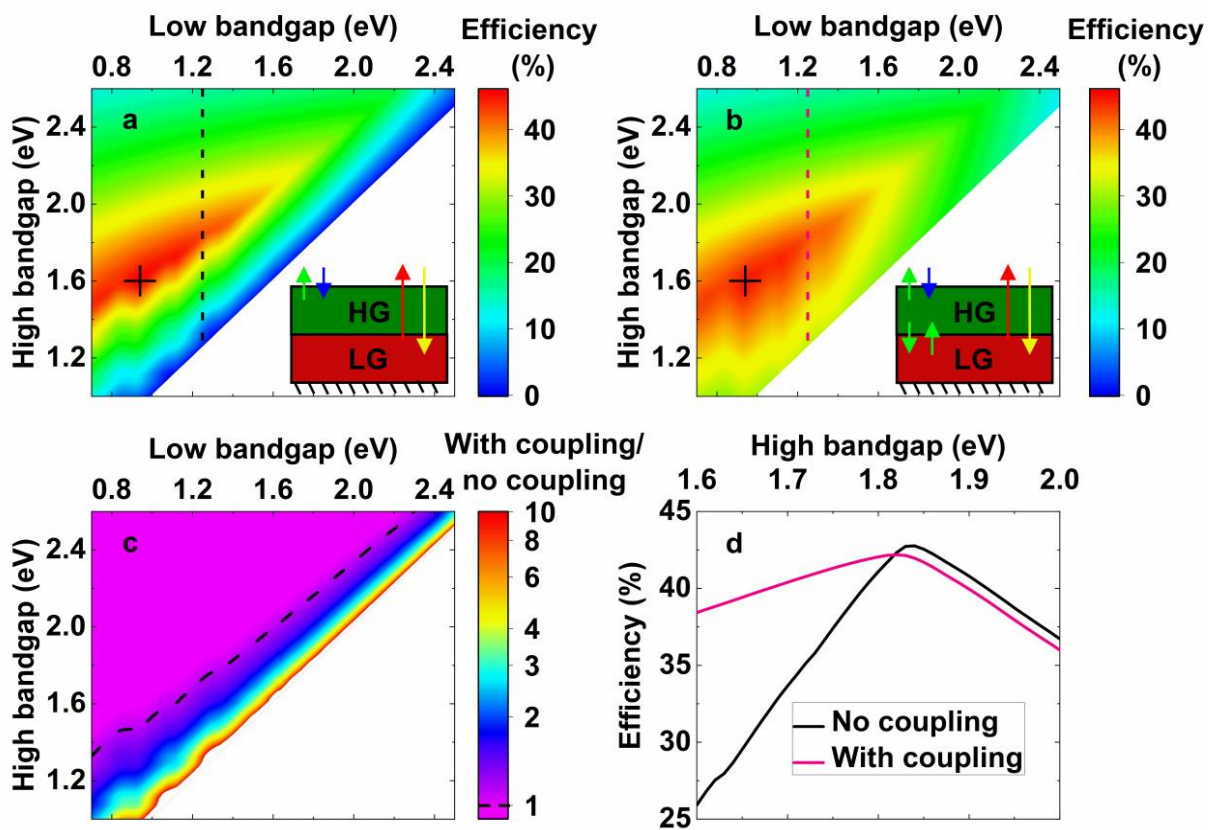

*Figure 1. The limiting efficiency of an ideal Shockley-Queisser-like tandem (where all light above the bandgap is absorbed) (a) without and (b) with luminescence coupling. Crosses mark bandgap pairs yielding the highest efficiency. The inset schematics demonstrate the system being modelled, with LG and HG corresponding to the low- and high-bandgap sub-cells, respectively, blue and yellow arrows denote absorbed incident solar radiation in the HG and LG cell, respectively, and other arrows correspond to re-emitted light. c) The ratio of these two graphs, with the dashed line marking the case where there is no change when including luminescence coupling. d) Line slices of a) and b) (as marked by dotted vertical lines on the respective panels) showing the efficiency of tandem cells, without and with luminescence coupling, when the bandgap of the low energy absorber is set to 1.25 eV.*

We now focus on the state-of-the-art experimental $FAPb_{0.5}Sn_{0.5}I_3$ and $FA_{0.7}Cs_{0.3}Pb(I_{0.7}Br_{0.3})_3$ tandem compositions as low- (1.25 eV) and high- (1.7 eV) bandgap sub-cells, respectively[14,31], which we solution process as thin films (see Methods). We use a combination of transient absorption spectroscopy (TA) and photoluminescence quantum efficiency (PLQE) to quantify decay rates in each material. In halide perovskites, charges have been observed to decay according to first, second and third order mechanisms[32,33], typically interpreted as non-radiative recombination via traps (with rate $a$), bimolecular recombination (with rate $b$, a component of which is radiative[32]) and non-radiative Auger recombination (rate $c$), respectively. These processes are described by:

$$\frac{dn}{dt} = -an - bn^2 - cn^3 \qquad (2)$$

where $n$ is the excited charge density and $t$ is time. In both materials we observe a broad ground state bleach in TA which scales linearly with excitation density (see SI Figure S2), and we integrate about the peak of this bleach. We estimate the excitation density, $n$, using the same approach as Richter et al.[32], and by scaling the bleach appropriately we present $\frac{dn}{dt}$ versus $n$ in Figure 2a for the FA$_{0.7}$Cs$_{0.3}$Pb(I$_{0.7}$Br$_{0.3}$)$_3$ thin film (see Methods). We fit this decay with the first, second and third order decay rates described in Equation 2 (red line). For our PLQE measurements, we consider

$$PLQE \times G_{ext} = \eta_{esc} b_r (p_i n + n^2), \qquad (3)$$

where $G_{ext}$ is the laser generation rate, $\eta_{esc}$ is the photon escape probability, $b_r$ is the internal radiative bimolecular recombination rate and $p_i$ is the background hole concentration (in the case of a p-type material). By measuring the laser generation rate, $G_{ext}$, and calculating $n$ from values obtained in our TA measurements (as $G_{ext} = an + bn^2 + cn^3$), we fit our PLQE data using Equation (3) to extract the background hole concentration and $\eta_{esc} b_r$ (Figure 2b). We note that we do not observe any phase segregation during these measurements (SI Figure S2). For the low-bandgap system, we use our previously reported doping densities, radiative bimolecular and Auger recombination rates (Table 1)[31].

To measure optical constants for both materials we combined ellipsometry and photothermal-deflection spectroscopy (PDS) measurements[19] (SI Figure S3). For the purposes of our calculations, we fit the below-bandgap region with an Urbach tail using photoluminescence (PL) spectra in order to only quantify absorption that clearly contributes to PL (SI Figure S4). The combination of these measurements and Urbach fit gives us absorption coefficients, $\alpha(E)$, and refractive indices, $n(E)$, for all relevant energies, which we plot in Figure 2c and 2d for the high- and low-bandgap systems, respectively. Using these optical constants, we simulate the internal PL spectra of our materials and compare these with the recorded (external) PL spectra to determine the escape probability, $\eta_{esc}$, for each sample without the need for any assumptions about the absorption of the material (see SI Figure S5 and SI Note 2 for further details). The value of $\eta_{esc}$ allows us to calculate the intrinsic radiative rate, $b_r$. We estimate the background minority carrier concentration, $n_i$ following the approach of Pazos-Outón et al.[19] and use our measurements values to calculate the equivalent intrinsic doping density, $\sqrt{n_i p_i}$. We note that although the low- and high-bandgap perovskites we measured are observed to be doped systems, we herein model the absorber layers as intrinsic layers because such background doping densities do not significantly affect limiting efficiencies (see SI Figure S6 and SI Note 3). We summarise all relevant experimentally extracted parameters for calculations in Table 1 and others in SI Table 1.

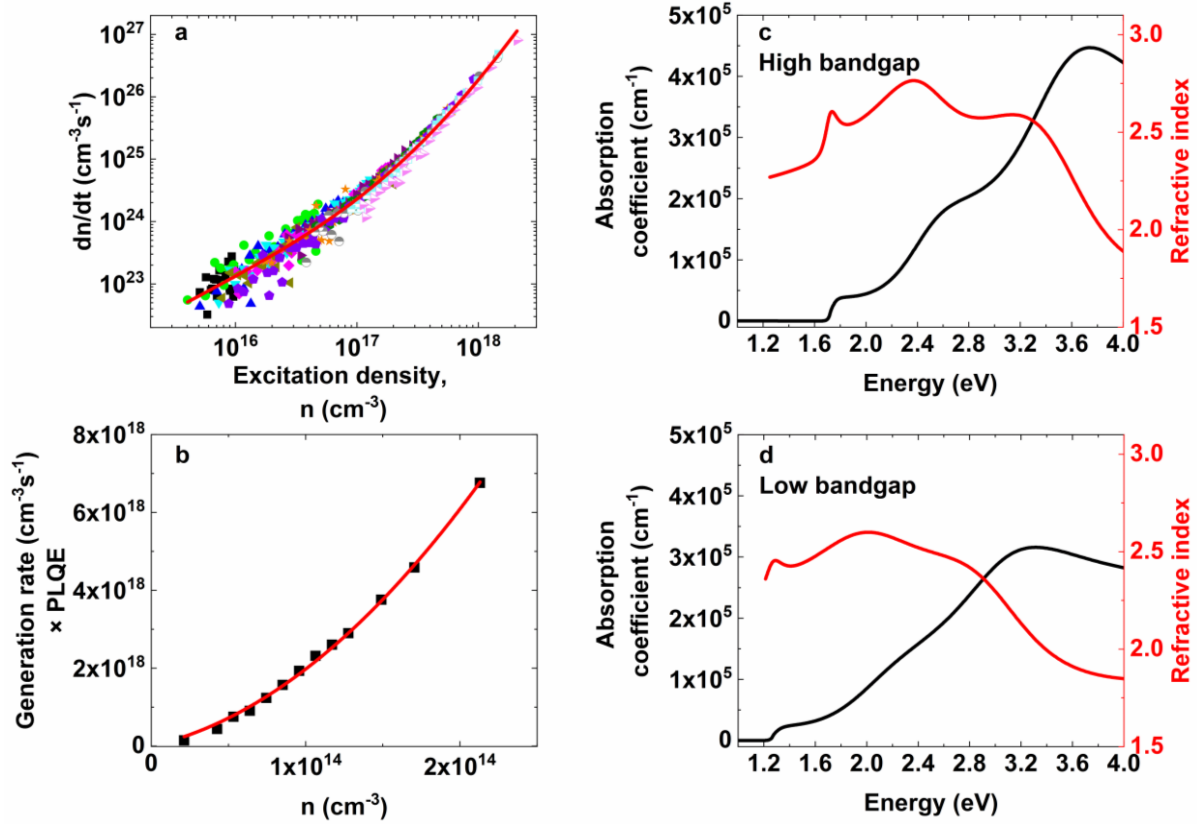

*Figure 2. a) Plot of dn/dt vs carrier density n (time is an implicit variable) extracted from transient absorption decay measurements of $FA_{0.7}Cs_{0.3}Pb(I_{0.7}Br_{0.3})_3$ thin films. Each symbol represents a different decay measurement, corresponding to a different initial excitation density. The red line is a fit to the data using Equation 2. b) Plot of generation rate × PLQE (black symbols) vs carrier density n (extracted from the TA measurements) for $FA_{0.7}Cs_{0.3}Pb(I_{0.7}Br_{0.3})_3$ thin films. The red line is a fit to the data using Equation 3. Absorption coefficients and refractive indices of c) $FA_{0.7}Cs_{0.3}Pb(I_{0.7}Br_{0.3})_3$ and d) $FAPb_{0.5}Sn_{0.5}I_3$ as measured by a combination of ellipsometry and photothermal deflection spectroscopy (see SI Figure S3 and SI Note 2 for details). Parameters extracted from these fits and optical analysis are summarised in Table 1 and SI Table 1.*

|  | $FAPb_{0.5}Sn_{0.5}I_3$ | $FA_{0.7}Cs_{0.3}Pb(I_{0.7}Br_{0.3})_3$ |
|---|---|---|
| *Internal bimolecular recombination rate, $b_r$ ($cm^3 s^{-1}$)* | $(3.0 \pm 0.2) \times 10^{-12}$ | $(5.1 \pm 0.2) \times 10^{-10}$ |
| *Auger recombination rate, $c$ ($cm^6 s^{-1}$)* | $(6.5 \pm 2.0) \times 10^{-29}$ | $(7.5 \pm 2.0) \times 10^{-29}$ |
| $n_i p_i$ ($cm^{-6}$) | $(2.7 \pm 0.2) \times 10^{17}$ | $(1.2 \pm 0.1) \times 10^8$ |
| *Urbach energy (meV)* | $16.1 \pm 0.1$ | $14.4 \pm 0.1$ |

*Table 1. Relevant parameters extracted from the time-resolved and steady-state optical characterisation of $FAPb_{0.5}Sn_{0.5}I_3$ and $FA_{0.7}Cs_{0.3}Pb(I_{0.7}Br_{0.3})_3$ thin films. See SI Table 1 for all extracted parameters from our measurements.*

A key ingredient in a limiting efficiency calculation is the fraction of sunlight absorbed at each energy $E$, $a(E)$, as calculated from the measured absorption coefficients and refractive indices (Figure 2). Yablonovitch demonstrated that it is possible to increase the absorption of a semiconductor significantly beyond an exponential Beer-Lambert type law close to its bandgap by considering rough front and back surfaces which randomise the direction of light inside a semiconductor, and a perfect back reflector[34]. This model, which we term Randomised, has previously been used as the workhorse for calculating the absorption of idealised single bandgap perovskite solar cells[19,35]. However, in a tandem stack the Randomised model predicts weak absorption above the bandgap in the low-bandgap sub-cell when compared to Beer-Lambert absorption, as is shown in Figure 3a (for a tandem stack of $FAPb_{0.5}Sn_{0.5}I_3$ and $FA_{0.7}Cs_{0.3}Pb(I_{0.7}Br_{0.3})_3$ assuming no parasitic absorption and sub-cell thicknesses of 1000 nm and 440 nm respectively, see SI Figure S7 for high bandgap results). This is due to light being treated as black-body radiation once it has entered the high-bandgap material, meaning some of it never reaches the low-bandgap sub-cell, but is instead directly re-emitted to the surroundings. To resolve this problem, we use a more advanced Lambertian absorption model which combines Randomised and Beer-Lambert type absorptances, termed Hybrid. This is an extension of Green's Lambertian absorptance model to idealised tandem solar cells[36] (see SI Notes 4 and 5 for full details). We apply this absorption model to single bandgap perovskite solar cells in SI Note 5 and demonstrate the limiting efficiency of low-bandgap $FAPb_{0.5}Sn_{0.5}I_3$ perovskite as 32.1 % (SI Figure S8). We note that photon recycling within a single perovskite layer is implicitly included within the Hybrid model.

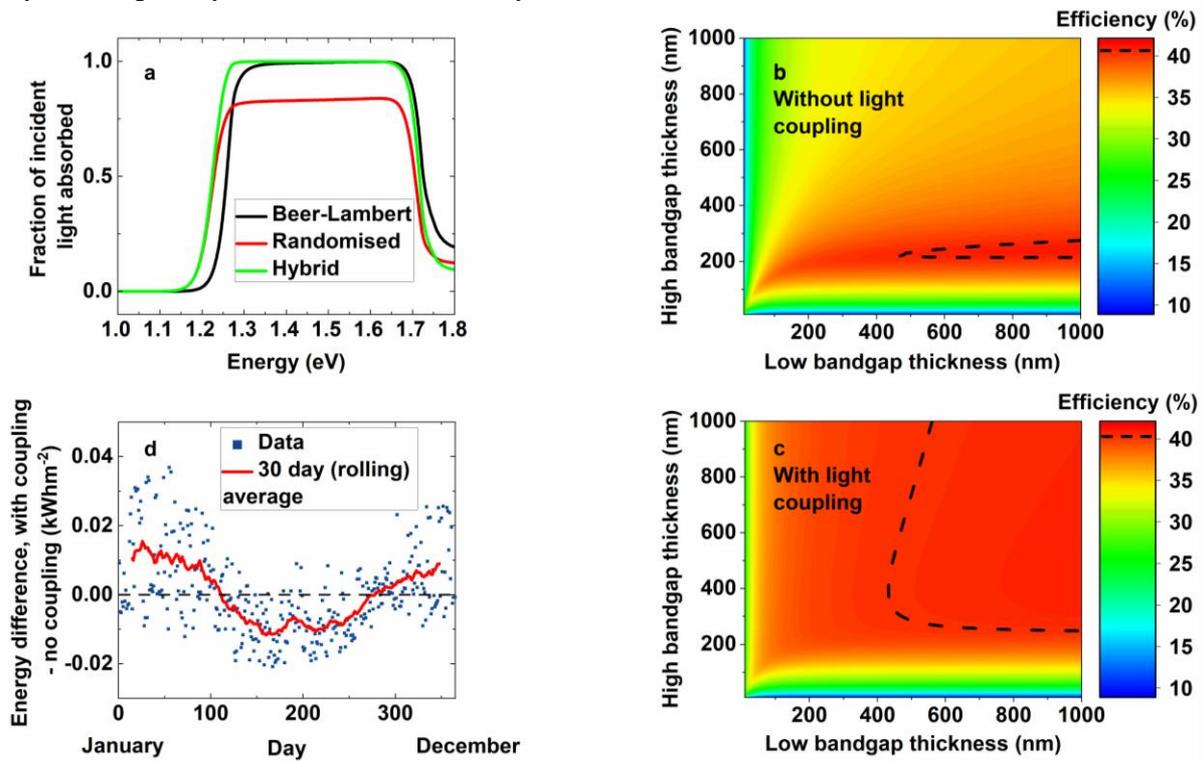

*Figure 3. a) Absorptance in the low-bandgap $FAPb_{0.5}Sn_{0.5}I_3$ absorber of the tandem solar cell stack, for the three absorptance models described in the main text, for low- and high-bandgap sub-cell thicknesses of 1000 nm and 440 nm respectively. We present the limiting efficiency of an all-perovskite ($FAPb_{0.5}Sn_{0.5}I_3$ and $FA_{0.7}Cs_{0.3}Pb(I_{0.7}Br_{0.3})_3$) tandem as a function of sub-cell thickness (b) without including and (c) including luminescence coupling in models, ascertained by using experimentally measured parameters but setting charge trapping to zero in both cases. The dashed lines denote regions within 1 % of maximum efficiency. d) The difference in energy generated with and without luminescence coupling throughout the year for North Roseau on the USA-Canada border. Sub-cell thicknesses are chosen to maximise energy yield in these simulations of 1000 nm and 220 nm without luminescence coupling/1000 nm and 310 nm with luminescence coupling, for low- and high-bandgap sub-cells, respectively (see SI Figure S12).*

To calculate the fundamental limiting efficiency of a solar cell at maximum power point, the absorption is maximised (using the Hybrid model, which gives less absorption than in Shockley-Queisser tandems) and the recombination rate is minimised by setting all controllable loss mechanisms to zero. Therefore, only radiative recombination and intrinsic non-radiative Auger recombination are included in our limiting efficiency calculations, while the effects of charge trapping are considered later. We also assume an equal Fermi-level splitting such that the populations of electrons ($n$) and holes ($p$) as a function of applied voltage $V$ follow $n = p = n_i e^{\frac{qV}{2k_BT}}$ in our intrinsic approximation, where $q$ is the charge of an electron and $k_B T$ the thermal energy.

The limiting efficiency of an all-perovskite tandem under AM1.5 as a function of sub-cell thicknesses with experimental-parameterised absorption and recombination coefficients, in the case where luminescence coupling is not considered in the modelling, is presented in Figure 3b (see SI Note 6 for full calculation details). We emphasise that removing light coupling is not possible in a real device, but this hypothetical case has been assumed in the literature to date and instructive to subsequently demonstrate the importance of this effect. We only carry out simulations to 1000 nm as diffusion limitations (which are not included in this model) are likely to become critical at higher thicknesses[19]. The maximum efficiency achievable is 41.1 % for optimal sub-cell thicknesses of 1000 nm (low-gap) and 240 nm (high-gap). To be within 1 % of the maximum efficiency the low- and high-bandgap thicknesses need to be in the range 570-1000 nm and 220-270 nm, respectively (dashed line on figure). This is a narrow range of thicknesses for the sub-cells due to the requirement for near-perfect short-circuit current matching, and imposes significant restrictions on device materials tunability. In Figure 3c, we present the limiting efficiency of the same system but now including the physically intrinsic process of luminescence coupling in our modelling (see details on modelling luminescence coupling between sub-cells in SI Note 7 and SI Figure S9). As in Figure 1, the maximum efficiency is slightly reduced in the presence of luminescence coupling (40.8 % for low- and high-bandgap sub-cell thicknesses of 1000 nm and 440 nm). For comparison, we note that the limiting efficiency of a Shockley-Queisser tandem (where all light is absorbed above the bandgap) with luminescence coupling is 44.9 % (cf. Figure 1b). In the system modelled here, to be within 1% of the maximum efficiency the thickness ranges are now 480-1000 nm and 260-1000 nm for the low- and high-bandgap sub-cells, respectively (dashed line on Figure 3c). This demonstrates a substantial increase in thickness tolerance due to any discrepancy in current matching being partly self-corrected through luminescence coupling, as has been discussed in idealised systems[22]. We note the combined thickness of both absorber layers can also be reduced by ~ 10 % when luminescence coupling is included, giving an additional advantage for lightweight applications. We present limiting efficiencies based on the Beer-Lambert model in SI Figure S10, where a limiting efficiency of 39.7 % is found when luminescence coupling is not considered, and 39.4 % when luminescence coupling is included.

We also model a perovskite-silicon tandem by coupling the same high-bandgap perovskite to an idealised silicon sub-cell (see SI Note 3 for details and SI Figure S11 for results). This perovskite has a bandgap better matched to that of silicon than of the low-bandgap perovskite considered above (c.f. Figure 1a), giving a limiting efficiency of 43.0 % without luminescent coupling being included in calculations (for sub-cell thicknesses of 580 μm and 1000 nm in the silicon and perovskite sub-cells, respectively). The efficiency limit is reduced to 42.0 % when the more physically realistic model, including luminescence coupling, is used for respective sub-cell thicknesses of 270 μm and 1000 nm. Luminescence coupling is again seen to increase the sub-cell thickness tolerance, in particular allowing for thinner low-bandgap (silicon) sub-cells, with the most commercially relevant silicon thicknesses of ~180 µm within 1 % of the maximum calculated efficiency. Furthermore, even thinner silicon sub-cells still give efficiencies close to the maximum (e.g. 50 µm is within 2 % of the maximum efficiency), which could allow for a range of ultrathin silicon fabrication techniques with possible cost benefits[37].

To explore how luminescence coupling affects tolerance to real-world spectra we again consider the all-perovskite tandem cells constructed from our experimental films and calculate the energy generated from a year's worth of irradiance spectra without including and including luminescence coupling. We used a typical meteorological years' worth of data from the National Solar Radiation Database, which includes spectrally resolved data and temperature variation, for a region on the border between United States and Canada (North Roseau) that represents reasonable spectral variation throughout the year[13,38]. We also note that the Lambertian absorption model treats incident light from all angles equally, allowing for a simplification in the calculations. We first calculate the total energy generated for a range of different sub-cell thicknesses, as presented in SI Figure S12a and b. The optimal thickness of the high-bandgap sub-cell is reduced compared to AM1.5 (240 nm to 220 nm without coupling, 440 nm to 310 nm with coupling), due to North Roseau having fewer clear days and thus less blue light than AM1.5. More importantly, while perovskite tandems including the intrinsic process of luminescence coupling in modelling gave a lower efficiency under AM1.5, a comparable total energy yield is generated over the course of a year (492.4 kWhm$^{-2}$ with luminescence coupling, compared to 492.1 kWhm$^{-2}$ without). This energy is also generated at different times of the year, as shown in Figure 3d, which shows the difference in energy generation with and without luminescence coupling each day for optimal thicknesses. When luminescence coupling is included in models more energy is generated in the winter months, while slightly less is generated mid-summer. We explain this by noting that the days in winter have a less blue spectrum (SI Figure S12c). If a solar cell is optimized for these less blue conditions, then in mid-summer (a bluer spectrum) luminescence coupling can transfer current from the high- to low-bandgap sub-cell, correcting for the mismatch in current. We confirm this by calculating the percentage of current from low-bandgap sub-cell that is generated from luminescence coupling, which is closer to zero in winter but increases up to 10 % in mid-summer (SI Figure S12d). These results demonstrate the increased spectral tolerance imparted on an all-perovskite tandem cell design when considering luminescence coupling in real world conditions, in agreement with previous analyses on idealised systems by Brown and Green[21].

In Figure 4a, we present the limiting efficiency of an experimentally parameterised all-perovskite tandem cell, including the intrinsic process of luminescence coupling in calculations, as a function of the non-radiative charge trapping rate $a$ for optimised thicknesses of 1000 nm and 440 nm. We observe that increasing the charge trapping rate in either material has a similar effect in terms of reducing the efficiency of the tandem. Figure 4b presents the ratio of Figure 4a to an equivalent calculation neglecting luminescence coupling in models (SI Figure S13). It is clear that luminescence coupling only plays a significant role when charge trapping rates $a < 10^6$ s$^{-1}$ in the high-bandgap sub-cell, equivalent to charge lifetimes being longer than 1 µs in the charge trapping regime. Furthermore, we confirm that even with non-zero charge trapping rates, current matching conditions are relaxed when luminescence coupling is included in simulations compared to the case that doesn't consider luminescence coupling (SI Figure S14). This critical charge trapping rate in the high-bandgap cell for luminescent coupling to be important in perovskite-silicon cells is ~10$^5$ s$^{-1}$ (SI Figure S11). We attribute this lower charge trapping rate in the perovskite-silicon tandems to the sub-cells having better bandgap matching, meaning that the low-bandgap sub-cell is not current-limiting, and thus charge densities (and hence likelihood of radiative recombination) are lower in the high-bandgap sub-cell. Other simulations confirm the trapping rate for luminescent coupling varies between $10^5$ s$^{-1}$ and $10^6$ s$^{-1}$ depending on how well current-matched the sub-cells are (see SI Figure S15). In both tandem technologies, our results show that luminescence coupling becomes important when the high-bandgap sub-cell has an external photoluminescence quantum efficiency (PLQE) of at least ~0.1 % at maximum power point (see SI Figures S11, S13 and S15 for PLQE calculations for each case and SI Note 6). We mark trapping rates from current state of the art films in the literature with a cross on Figure 4a and 4b to demonstrate that we are already realising conditions in which luminescence coupling becomes important and thus we expect that these effects must be considered in further development of all tandem cells[14,31,39]. In order to maximise the luminescence coupling in a real tandem cell, any inter-layer between the sub-cells should have a (real) refractive index at least as high as the perovskite sub-cells, so the escape cone from high to low sub-cell remains as large as possible. We also emphasise that luminescence coupling is an intrinsic process which cannot be prevented from occurring (see SI Note 8).

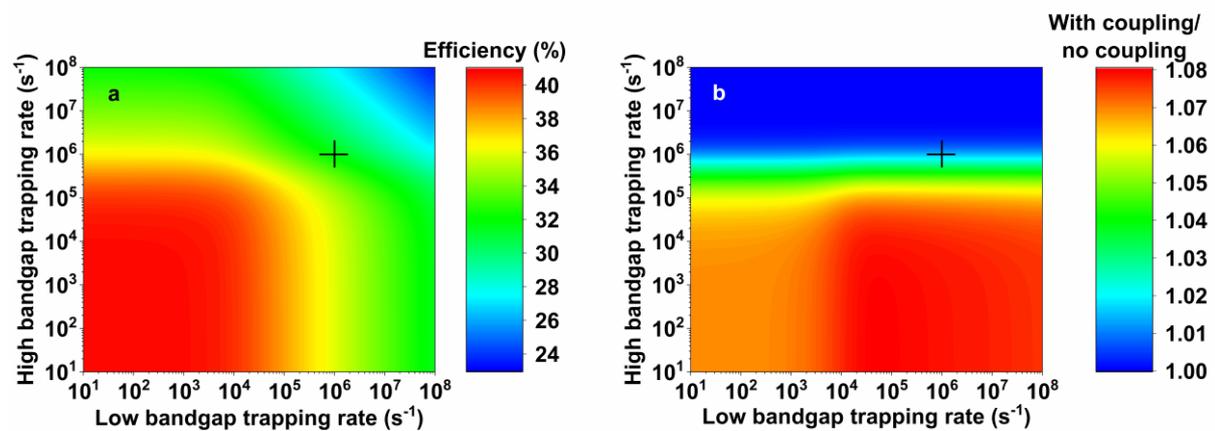

*Figure 4. Limiting efficiency of all-perovskite tandem solar cells comprised of the experimental FAPb$_{0.5}$Sn$_{0.5}$I$_3$ and FA$_{0.7}$Cs$_{0.3}$Pb(I$_{0.7}$Br$_{0.3}$)$_3$ absorbers with optimised thicknesses of 1000 nm and 440 nm as a function of charge trapping rate a) with luminescence coupling and b) the ratio of this model to the efficiency without including luminescence coupling for the same film thicknesses. Marked crosses correspond to charge trapping rates in current state of the art films[14,31,39].*

In order to experimentally demonstrate luminescence coupling and its effect on actual tandem devices, we perform measurements on an all-perovskite tandem cell following the device architecture of Palmstrom et al.[40]. We first consider a case where the high-bandgap sub-cell has significantly higher short-circuit current than the low-bandgap sub-cell through selective illumination of the top cell with 405 nm excitation (absorption depth < 50 nm). We observe current from the device at different applied voltages (SI Figure S16) and, importantly, luminescence from the high-bandgap sub-cell at all applied voltages (Figure 5a); even when the tandem stack is at short-circuit, the high-bandgap luminescence is still 4 % of the intensity as at open-circuit. Furthermore, we determine the quasi-fermi-level-splitting (QFLS) of the high- and low-bandgap sub-cells by analysing the photoluminescence (PL) properties of each absorber layer in a device stack (see methods and SI Figure S17 for fits), which corresponds to the maximum open circuit voltage ($V_{OC}$) that each sub-cell can contribute to the tandem stack. Under 405-nm excitation, we observe that the $V_{OC}$ of the tandem exceeds the QFLS of the high-bandgap sub-cell, meaning that the low-bandgap sub-cell must be contributing notable voltage, despite the high-bandgap absorbing nearly all photons (SI Figure S18).

A confocal PL map of a cross-section of the tandem using appropriate optical filters to selectively observe emission from the high- or low- bandgap sub-cell is shown in Figure 5b. We then excited the centre of the high-bandgap sub-cell with a pulsed excitation and, while keeping the excitation spot fixed, spatially scanned the selective PL detection across the cross-section (Figure 5c), revealing emission from the low-bandgap cell after excitation in the high-gap cell. To confirm this is luminescence coupling, we consider that the number of photons absorbed in the low-gap at time $t$ is proportional to the time-resolved photoluminescence (TRPL) from the high-gap, $PL_{HG}(t)$. At low excitation densities the low-gap TRPL is extremely short (SI Fig S19). Therefore, if the TRPL signal from the low-gap absorber at time $t$ is due to recombination of excited electrons and holes, the quantity should be proportional to $PL_{HG}(t)^2$; this is exactly what we observe in Figure 5d (see SI Note 9 for further discussion and additional cross section results). These collective results demonstrate that the high-bandgap sub-cell is luminescent within an operating tandem stack, and that these emitted photons can be absorbed in the low-bandgap sub-cell (i.e. luminescence coupling).

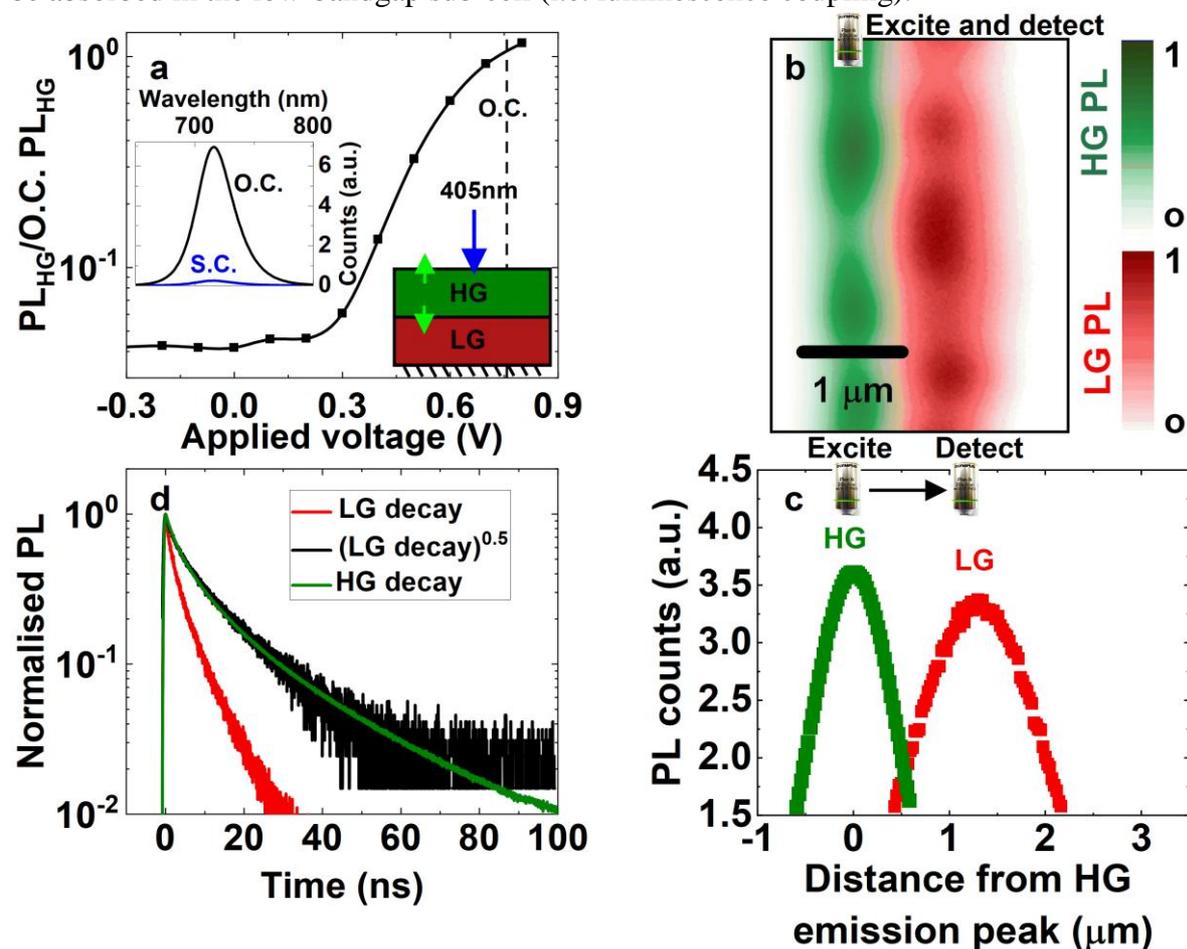

*Figure 5. a) Photoluminescence (PL) of the high-bandgap (HG) sub-cell in the tandem stack, relative to that when the tandem is held at open circuit (O.C.), as a function of voltage when the HG sub-cell is selectively excited with 405 nm excitation from the top side, as shown in schematic (LG corresponds to the low-bandgap sub-cell). Inset graph shows the HG PL when the tandem stack is held at O.C. and short-circuit (S.C.). b) PL maps of the tandem cross section when exciting with a 636 nm laser and using a 775 nm short pass or long pass filter to only observe PL from the high- or low-bandgap sub-cells, respectively. c) Fixed excitation at the centre of the HG sub-cell with 636 nm excitation and spatially varying the PL detection away from the excitation spot across the device cross-section, using 750-nm short pass and 800-nm long pass filters to collect emission from HG and LG materials, respectively. Note distance*

*scale here does not exactly correspond to distances on the sample surface (see SI Note 9). d) Time-resolved PL of the HG and LG regions, as well as the square root of the low-bandgap decay to show the match with the HG decay.*

**Conclusion**

In this work we have calculated the limiting efficiencies of perovskite-based tandem solar cells, including all intrinsic loss processes and luminescence coupling between sub-cells. By measuring the recombination rates and absorption coefficients of low- and high-bandgap perovskite films, we calculated the limiting efficiency of an all-perovskite tandem as 40.8 % and a perovskite-silicon tandem as 42.0 % when the intrinsic process of luminescence coupling between the sub-cells is included and in the absence of trapping. We show that current state of the art high-bandgap perovskite films for tandem cells have charge trapping rates and luminescence quantum efficiencies on the order required for luminescence coupling to play an important role in devices, which reduces the need for short-circuit current matching compared to earlier predictions that do not consider luminescence coupling. We demonstrate luminescence coupling in perovskite tandems, when compared to a model which ignores this effect, gives increased tolerance in choice of bandgaps, sub-cell thicknesses, and greater tolerance to a range of real-world spectra, and hence relaxes the previously determined criteria for materials and device design. We conclude with a new design rule for perovskite tandems: it is always better for the high-bandgap material to have the higher short circuit current, as any discrepancy in current matching will be partially corrected by luminescence coupling between sub-cells. Importantly, these guidelines allow unstable >1.7 eV high-bandgap perovskite absorbers to be avoided when targeting maximum performance. We also present experimental evidence of luminescence coupling occurring in an all-perovskite tandem including visualisation of the effect, highlighting the importance of the effect in ongoing perovskite tandem developments.

**Methods**

*Film and device preparation details*

Low-bandgap $FAPb_{0.5}Sn_{0.5}I_3$ samples were prepared in a nitrogen-filled glovebox as described in our previous work[31]. We fabricated samples with 5 % Zinc content as these were the most air stable without affecting photoluminescence or absorption properties.

To prepare the high-bandgap perovskite, a 1.2 M solution of $FA_{0.7}Cs_{0.3}PbI_{0.7}Br_{0.3}$ was used. First, 304.3 mg $PbI_2$ (TCI) and 198.2 mg $PbBr_2$ (TCI) were dissolved in 0.8 mL anhydrous N,N-dimethylformamide (sigma) and the solution heated to 100 °C. The solution was then cooled to room temperature and 144.4 mg FAI (greatcell solar) was added. For the CsI (sigma) precursor, we prepared 1.8 M CsI solution in dimethyl sulfoxide (sigma) and heated the solution to 150 °C. Finally, the FAPb(IBr) and CsI solutions were mixed with a 4:1 volume ratio to achieve the desired composition to finalise the precursor solutions. Prior to spin-coating, the glovebox was purged for 15 minutes with nitrogen to achieve a clean atmosphere. To prepare the films, 50 µl of perovskite precursor solution was deposited on a glass, quartz or silicon substrate (which had been cleaned by sonication in anisole and isopropanol) and spin-coated at 1000 rpm for 10 s and then 6000 rpm for 20s. 100 µl of chlorobenzene anti-solvent was dynamically dropped on the middle of substrate, 5 seconds before the end of the spinning protocol. The films were then annealed on a hotplate at 100 °C for 30 minutes in the glove box. The spin-coating method is optimised for an integrated spin-coater in an MBRAUN glovebox with the lid open. To confirm phase purity we carried our X-Ray diffraction on our high-bandgap samples and results are shown in SI Figure S20.

All films for photoluminescence quantum efficiency and transient absorption spectroscopy measurements were encapsulated with a glass cover slip using a transparent UV epoxy (Blufixx).

Perovskite-perovskite solar cells were fabricated following the procedures of Palmstrom et al.[40]. The champion perovskite/perovskite tandem solar cell exhibited efficiency of 23.1 %, short circuit current of 16.0 mAcm$^{-2}$ and open-circuit voltage 1.88 V. Owing to the multiple spin-coated layers and many involved steps in heavily used multi-user facilities the performance distribution however is wide and the average efficiency amounted to around 16 %. For the optical characterizations presented herein we further prepared all perovskite tandem solar cells on quartz substrates with a home-made ITO coating to facilitate improved excitation at shorter wavelengths, which yielded in a lower device yield in comparison to commercial ITO on glass substrates. As shown below, the device under test exhibited a stabilised power conversion efficiency of 13.2 %, an open circuit voltage of 1.66 V and a short circuit current of 14 mAcm$^{-2}$ (see device characteristics in SI Figure S21). Measurements were performed on encapsulated devices (unless stated otherwise) at different stages over a period of 12 months, with storage in a nitrogen glove box between measurements.

*Photoluminescence and photoluminescence quantum efficiency*

Photoluminescence quantum efficiency (PLQE) measurements were recorded using an integrating sphere, following the three measurement approach of De Mello et al.[41]. In both photoluminescence and PLQE measurements continuous wave temperature controlled Thorlabs 405 nm or 520 nm laser was used to photo-excite samples and excitation fluence varied with an optical filter wheel. The emission was recorded using an Andor IDus DU420A Silicon detector. Spot size was recorded using a Thorlabs beam profiler, where the size was set to be to where the intensity of the beam falls to $1/e^2$. Voltages were applied to and currents recorded from the tandem with a 2600 series Keithley source-meter.

*Transient absorption spectroscopy*

For the pump, sub-ns pulses at 532 nm were generated by a Picolo-25 MOPA laser (InnoLas). A Spectra Physics Solstice Ti:Sapphire laser generated 90 fs pulses at a frequency of 1 kHz and a broad band probe beam was generated using a home-built noncollinear optical parametric amplifier. Probe and reference beams were measured with a Si dual-line array detector (Hamamatsu S8381-1024Q) and read by a board from Stresing Entwicklungbüro (custom made). Excitation fluence varied with an optical filter wheel. Other details are similar to those in Rao et al.[42]. Spot size was recorded using a Thorlabs beam profiler, where the size was set to be where the intensity of the beam falls to $1/e^2$.

*Photothermal deflection optical absorption spectroscopy*

A monochromatic pump light beam was impinged on the sample (film on quartz substrate), which on absorption produced a thermal gradient near the sample surface via non-radiative relaxation induced heating. This results in a refractive index gradient in the area surrounding the sample surface. This refractive index gradient was further enhanced by immersing the sample in an inert liquid FC-72 Fluorinert® (3M Company) which has a high refractive index change per unit change in temperature. A fixed wavelength continuous wave laser probe beam was passed through this refractive index gradient producing a deflection proportional to the absorbed light at that particular wavelength, which is detected by a photo-diode and lock-in amplifier combination. All samples were loaded into the inert liquid in a nitrogen filled glovebox to prevent effects of air or moisture exposure.

*Ellipsometry*

The complex refractive index of the different perovskite films, $\sqrt{\epsilon_b(E)} = n(E) - ik(E)$, was obtained from ellipsometry measurements at room temperature (Sopra PS-1000 SAM). We employed a genetic algorithm to fit the experimental data assuming a Forouhi-Bloomer model for three oscillators[43,44]:

$$n(E) = n_\infty + \sum_{j=1}^{3} \frac{B(E - E_j) + C_j}{(E - E_j)^2 + \Gamma_j^2}; \quad k(\omega) = \begin{cases} \sum_{j=1}^{3} \frac{f_i(E - E_j)^2}{(E - E_j)^2 + \Gamma_j^2}, & E > E_g \\ 0, & E \leq E_g \end{cases}$$

and

$$B_j = \frac{f_j}{\Gamma_j}\left(\Gamma_j^2 - (E - E_j)^2\right) \text{ and } C_j = 2f_i\Gamma_j(E - E_j).$$

Here $E$ is the photon energy, $E_g$ is the bandgap energy, and $E_j$, $f_j$, and $\Gamma_j$ are the position, strength, and width of one oscillator. In order to make the method more accurate and reliable, three different incident angle (60º, 65º and 70º) measurements were fitted simultaneously. A layer of arbitrary thickness ($d_p$, 0 nm - 30 nm) and dielectric function $\varepsilon_{\text{eff}}$ models the surface roughness of the film assuming a Bruggeman effective medium approximation[45]:

$$\frac{1 - \epsilon_{\text{eff}}(E)}{1 + 2\epsilon_{\text{eff}}(E)} p_p + \frac{\epsilon_b(E) - \epsilon_{\text{eff}}(E)}{\epsilon_b(E) + 2\epsilon_{\text{eff}}(E)} (1 - p_p) = 0$$

where $p_p$ is the porosity, i.e. the air to perovskite ratio, of the layer. To reduce the impact of randomness on the initial fitting parameters, some of the parameters were appropriately bounded according to the experimental values of the layer thickness (estimated from cross-section scanning electron microscopy images) and the bandgap (extracted from the photoluminescence maximum). Fits to recorded data are presented in SI Figure S22 and results summarised in SI Table 2.

*Hyperspectral Photoluminescence Measurements (and definition of 1 sun for current-voltage measurements under 405 nm excitation)*

Absolute photoluminescence maps were recorded using a hyperspectral widefield imager from Photon etc. using a 20x objective. Following literature[46], the setup was calibrated in a two-step process: first using a calibrated halogen lamp that is coupled into an integrating sphere to obtain a spectral calibration and, secondly, using a fiber-coupled laser to perform an absolute calibration at one wavelength. All samples were excited with a 405 nm laser. For a 1 sun equivalent excitation density in the high gap, the laser intensity was set to 117 mWcm$^{-2}$, which corresponds to $2.4 \cdot 10^{17}$ photons cm$^{-2}$s$^{-1}$. Taking into account parasitic absorption of the laser excitation in the glass substrate and interlayers, of about 50%, $1.2 \cdot 10^{17}$ cm$^{-2}$s$^{-1}$ charge carriers are generated within the high and low gap absorbers. This value corresponds well to the generation rate of the individual sub-cells in a realistic tandem. Higher and lower excitations between 0.2 and 10 suns were calculated similarly (and also used in current-voltage measurements under 405 nm excitation, as in SI Figure S18).

*Calculation of QFLS*

The QFLS was calculated by assuming Lambertian emission via Würfel's generalized Planck law[47], which relates the spontaneous emission of photons in a direct semiconductor to the chemical potential of the non-equilibrium charge carrier concentration to the local temperature T providing that the specific absorptivity α(E) is known.

$$I_{PL}(E) = \frac{2\pi E^2 a(E)}{h^3 c^2} \cdot \frac{1}{\exp\left(\frac{E - QFLS}{k_B T} - 1\right)}$$

This equation can be simplified by assuming that the spectral absorptivity approaches unity for photon energies above the bandgap[48].

$$\ln\left(\frac{I_{PL}(E) h^3 c^2}{2\pi E^2}\right) = -\frac{E}{k_B T} + \frac{QFLS}{k_B T}$$

Here $I_{PL}$ is the measured absolute photoluminescence, E the photon energy, $k_B$ the Boltzmann constant, T the temperature, c the speed of light, h the Planck constant. By plotting the normalized PL spectra, we find that the local charge carrier temperature does not vary with excitation fluence within the here considered range. Therefore we fit the above equation to the high-energy slope of the PL emission, with fixed local charge carrier temperature to extract the QFLS. Example QFLS fittings are shown in SI Figure S17.

*Confocal time-resolved Photoluminescence Measurements*

To record time resolved photoluminescence and cross-sectional photoluminescence maps of the perovskite tandem solar cells, we used a confocal single-photon counting fluorescence microscope from Picoquant. In all cases, excitation was performed at 636 nm using a 100x long working distance air objective (NA = 0.8). The photoluminescence was collected through a dichroic mirror, a 640 nm long-pass filter, and a 50 µm pinhole onto a single photon counting SPAD detector. Typically, we raster scan both the excitation and detection using a galvo mirror system, where both the objective and sample remain at a fixed position. To observe the high- and low-bandgap films in cross sections (Figure 5b) we used a 5 MHz repetition rate and pulse energies of 1.8 µJ cm$^{-2}$ and 16.2 µJ cm$^{-2}$ respectively.

To investigate the light coupling between the high and low gap sub-cells in the perovskite tandem, we further fixed the excitation (within the high-gap subcell, at 1 MHz or 5 MHz repetition rate and pulse energy of 5.1 µJ cm$^{-2}$), and raster scanned the detection only using a galvano mirror system (scanning across the full cross section and averaging vertically over 0.5 µm in each case). To selectively detect emission from the high and low gap absorbers, we used additional 750 nm short-pass or 750 nm long pass filters respectively. Further discussion and interpretation of results is shown in SI Note 9. Lastly, we also observed diffraction features related to the excitation in and/or photoluminescence from both sub-cells at large distances, which was not explored further here.

*Preparation of cross-sections*

The cross-sectional PL mapping experiments require a well prepared cross-section. We obtained best results by: i) removing the standard encapsulation slide used to protect perovskite-perovskite tandem solar cells from humidity and oxygen ingress; ii) breaking the tandem solar cell with help of a scratch outside the area of interest; and iii) encapsulation of the cross-section 90° on an ultrathin microscopy cover slip. This allowed the acquisition of high-resolution PL maps without oxygen related degradation at the cross-section with a long-working objective.

*Atomic force microscopy*

Sample thickness was recorded using an Asylum Research MFP-3D atomic force microscope in non-contact AC mode. A scratch on the surface on an unencapsulated sample was made using metal tweezers and the average difference in height between the material surface and the glass below as recorded (after 0$^{th}$ order flattening and 1$^{st}$ order plane fit were applied). All measurements and data processing were carried out on Asylum Research AFM Software version 15.

*X-Ray Diffraction*

X-Ray diffraction was performed using a Bruker X-ray D8 Advance diffractometer with Cu Kα1,2 radiation (λ= 1.541 Å). Spectra were collected with an angular range of 5° < 2θ < 35°, Δθ = 0.15° with each measurement being recorded for 0.1 s.

*Simulations*

All simulations were carried out on home-built software, run in the programming language MATLAB.

**Acknowledgements**
ARB acknowledges funding from a Winton Studentship, Oppenheimer Studentship the Engineering and Physical Sciences Research Council (EPSRC) Doctoral Training Centre in Photovoltaics (CDT-PV). ARB thanks Luis Pazos-Outón for supplying data for MAPbI$_3$ solar cells. FL acknowledges financial support from the Alexander Von Humboldt Foundation via the Feodor Lynen program and thanks Prof. Sir R. Friend for supporting his Fellowship at the Cavendish Laboratory. Y-HC acknowledges the funding from Taiwan Cambridge Scholarship. AJ-S gratefully acknowledges a postdoctoral scholarship from the Max Planck Society. KF acknowledges a George and Lilian Schiff Studentship, Winton Studentship, the Engineering and Physical Sciences Research Council (EPSRC) studentship, Cambridge Trust Scholarship, and Robert Gardiner Scholarship. GE was funded by NREL's LDRD program. ER acknowledges the European Research Council (ERC) under the European Union's Horizon 2020 research and innovation programme (HYPERION, Grant Agreement Number 756962) and the EPSRC for a DTP Part Studentship. MA-J acknowledges funding support from EPSRC through the program grant: EP/M005143/1. MA-J thanks Cambridge Materials Limited for their funding and technical support. MA acknowledges funding from the European Research Council (ERC) (grant agreement No. 756962 [HYPERION]) and the Marie Skłodowska-Curie actions (grant agreement No. 841386) under the European Union's Horizon 2020 research and innovation programme. BVL acknowledges funding from the Max Planck Society, the Cluster of Excellence e-conversion and the Center for Nanoscience (CeNS). SDS acknowledges the Royal Society and Tata Group (UF150033) and the EPSRC (EP/R023980/1, EP/T02030X/1, EP/S030638/1). We thank Axel Palmstrom and William Nemeth at NREL for depositing some of the layers in the tandem stack.

**Conflicts of Interest**
SDS and GEE are Co-Founders of Swift Solar Inc.

**Data availability**
The data underlying this manuscript are available at [url to be added in proof].

**Code availability**
The codes used in this work are available at [url to be added in proof].